\documentclass{ws-mpla}

\begin{document}

\markboth{G.~Giardina {\it et al.}}
{Bremsstrahlung emission in $\alpha$-decay of $^{226}\mbox{\rm Ra}$}

\catchline{}{}{}{}{}

\title{Bremsstrahlung emission during $\alpha$-decay of $^{226}\mbox{\rm Ra}$}

\author{\footnotesize  GIORGIO GIARDINA, GIOVANNI FAZIO,
GIUSEPPE MANDAGLIO, \\ MARINA MANGANARO}

\address{Instituto Nazionale di Fisica Nucleare, Sezione di Catania,
and Dipartimento di Fisica dell'Universit{\`a} di Messina, 98166, Messina, Italy\\
giardina@nucleo.unime.it}

\author{SERGHEI~P.~MAYDANYUK, VLADISLAV~S.~OLKHOVSKY}

\address{Institute for Nuclear Research, National Academy of Science of Ukraine, 03680 Kiev, Ukraine}

\author{NIKOLAY V.~EREMIN, ANTON A.~PASKHALOV, DMITRY A.~SMIRNOV}

\address{Skobeltsin Institute of Nuclear Physics, Lomonosov Moscow State University, 119899 Moscow, Russia}

\author{CARMELO SACC\'{A}}

\address{Dipartimento di Scienze della Terra dell'Universit\'{a} di Messina, 98166 Messina,Italy}

\maketitle

\pub{Received (Day Month Year)}{Revised (Day Month Year)}

\begin{abstract}
We obtained the spectrum of probability  of the bremsstrahlung emission accompanying the $\alpha$-decay of $^{226}{\rm Ra}$ (E$_{\alpha}$=4.8 MeV) by measuring the $\alpha$-$\gamma$ coincidences  and using the model presented in our previous study on the $\alpha-$decay of  $^{214}{\rm Po}$ (E$_{\alpha}$=7.7 MeV). We compare the experimental data with the quantum mechanical calculation and find a good agreement between theory and experiment. We discuss the differences between the photon spectra connected with the $\alpha$-decay of the $^{226}{\rm Ra}$ and $^{214}{\rm Po}$ nuclei. For the two mentioned nuclei we analyze the bremsstrahlung emission contributions from the tunneling and external regions of the nucleus barrier into the total spectrum, and we find the destructive interference between these contributions.
We also find that the emission of photons during tunneling of the $\alpha$-particle gives an important contribution to the bremsstrahlung spectrum in the whole E$_{\gamma}$ energy range of the studied $^{226}$Ra nucleus. 

\keywords{Alpha-decay; photon bremsstrahlung; spectrum for $^{226}\mbox{Ra}$; sub-barrier and external contributions; interference; tunneling.}
\end{abstract}

\ccode{PACS Nos.: 23.60.+e; 23.20.Js; 41.60.-m; 03.65.XP; 27.90.+b.}

\section{Introduction}	

In recent years many experimental and theoretical efforts$^{1-12}$
have been made to investigate on the nature of the bremsstrahlung
emission in the $\alpha$-decay of heavy nuclei, because the
behavior of the energy spectrum of photons is strongly related to
the dynamics of the $\alpha$-decay and alpha--nucleus particle
potential. In some case the energy spectrum of bremsstrahlung
shows some slight oscillations\cite{EPJA_2008},
in other case authors observed a minimum\cite{Kasagi97,KasagiJPG},
in some experiment authors\cite{Boie.2007}
have not observed evidence of any structure. Therefore the main
problem to investigate the tunneling of the $\alpha$-particle
 through the Coulomb barrier of nucleus is to enlarge the
area of the study on other nuclei and to compare the
characteristics of the $\gamma$-spectra. In this paper we present
the results of the last experiment on the study of the
bremsstrahlung emission in $\alpha$-decay of $^{226}{\rm Ra}$. We
also analyze and discuss the comparison between the experimental and theoretical
results of the photon emission related to the $^{226}{\rm Ra}$ and $^{214}{\rm Po}$ nuclei. In Sec. 2 we shortly describe the characteristics of the experiment and the results of the  bremsstrahlung emission probability. In Sec. \ref{sec.3} we present the basis of the model, the analysis of our calculation about the various contributions of the bremsstrahlung emission, and we discuss the results of the $\gamma$-emission for the  $\alpha$-decay of $^{226}{\rm Ra}$, in comparison with the ones previously obtained for $^{214}{\rm Po}$. Sec. 4 is devoted to our conclusion.

\section{Experiment and results}

The experimental set up was the same as described in our previous
paper\cite{EPJA_2008}. The source of $^{226}{\rm Ra}$ with a
activity of about $10^{4}$  $\alpha$-particles$/$s  was used. Along the decay chain of this
nucleus the $\alpha$-particles were recorded. The diameter of the radioactive spot on
the source surface was about 8 mm and $\alpha$-particles were
detected by a silicon surface-barrier detector with energy
resolution of about 20 keV at the $\alpha$-particle energy of 4.8
MeV. The $\alpha$-detector was 200 mm$^{2}$ in area and was placed
at the distance  of about 1 cm from the source. The time
resolution of the $\alpha$-$\gamma$ coincidence technique was
$\tau$= 10 ns.

The $\gamma$-rays were detected by the NaI(Tl)-detector with diameter
of 3 cm and thickness of 3 cm, and the distance between source and
$\gamma$-detector has been about 1.6 cm. The angles between the
two detectors and the normal axis to the surface of the source
were $45^{\circ}$. So the total angle between the $\alpha$ and
$\gamma$ detectors was about $90^{\circ}$. This angle value was
chosen to increase the yield of E1 dipole bremsstrahlung photons
and to reduce the influence of $\gamma$-rays from E2-quadrupole
transitions of excited states of daughter nuclei.

The absolute values of the NaI(Tl)-detector efficiency  was determined by measuring the
intensities of lines of the standard $\gamma$-sources $^{241}{\rm
Am}$ (E$_{\gamma}$=59.6 keV), $^{57}{\rm Co}$ (122 and 136 keV),
$^{226}{\rm Ra}$ (186, 295, 532 and 609.4 keV), $^{137}{\rm Cs}$
(662 keV) and $^{60}{\rm Co}$ (1.17 and 1.33 MeV), by replacing
each of them in the position of the $^{226}{\rm Ra}$-source of
$\alpha$-particles.The measurements of bremsstrahlung photons in coincidence with $\alpha$-particles for  five $\alpha$-groups\cite{Firestone96} from decays of $^{226}{\rm Ra}$ (E$_{\alpha}$= 4.784 MeV), $^{210}{\rm Po}$
(5.304 MeV), $^{222}{\rm Rn}$ (5.490 MeV), $^{218}{\rm Po}$ (6.002
MeV) and $^{214}{\rm Po}$ (7.687 MeV) were
performed during about 1500 hours and the total number of 4.4$\cdot$10$^{6}$ events was registered by the  $\alpha$-$\gamma$ coincidence system.   The analysis of these events in the
(E$_{\gamma}$ vs E$_{\alpha}$)-plane located in the region of the total
energy conservation line E$_{\gamma}$ + E$_{\alpha}$=
\textit{constant}, taking into account the detectors energy
resolutions (20 keV for $\alpha$-particles and 32 keV for
photons), gives us the possibility to determine the yield of
photons at energies up to 803 keV (1-st excited state of the
daughter $^{206}$Pb nucleus) for the $\alpha$-decay of $^{210}{\rm
Po}$. The range of (E$_{\gamma}$ vs E$_{\alpha}$)-coincidences was
collected for one of the experimental run with the measurement
time of about 150 hours.

For other details on the recorded events and analysis, see our previous paper \cite{EPJA_2008} and references therein.

To take into account the defined angular dimensions of detectors
we use the same procedure to obtain the angular averaged
probability of the photon emission dP/dE$_{\gamma}$ as described
in Refs.~\refcite{Kasagi97,KasagiJPG}, where the $\alpha-\gamma$ angular
correlation function W($\theta$), for the case of E1-dipole photon
emission and point-like $\alpha-\gamma$ source is
\begin{equation}
W(\theta)=1+A_{2}\cdot Q_{2}\cdot P_{2}(cos\theta)\hspace{1.5cm}
\end{equation}
where  A$_{2}$= -1 for the dipole E1 transitions,
P$_{2}$(cos$\theta$) is the second order polynome of Legendre,
Q$_{2}$ is the total geometrical attenuation coefficient of the
second order: Q$_{2}$= Q$_{2}^{\alpha}$ $\cdot$ Q$_{2}^{\gamma}$,
Q$_{2}^{\alpha,\gamma}$ being the geometrical attenuation
coefficients for $\alpha$-particle and photon detectors.

The exact calculations in the framework of the density matrix
theory\cite{Ferguson69,Camp69} give us the real experimental geometry
values of the attenuation coefficients: for the $\alpha$-particle
detector it is Q$_{2}^{\alpha}$=0.90, and for the photon detector
the value of Q$_{2}^{\gamma}$-coefficent is varied from
Q$_{2}^{\gamma}$= 0.66 for E$_{\gamma}$= 100 keV up to
Q$_{2}^{\gamma}$= 0.76 for E$_{\gamma}$= 800 keV.

The probability of the photon emission at angle $\theta$ can be
written as\cite{D'arrigo94,Eremin00}
\begin{equation}
\frac{d^{2}P}{dE_{\gamma}d\Omega_{\gamma}}=N_{\alpha-\gamma}(\theta,E_{\gamma})/(\Delta
T_{meas}\cdot n_{\alpha}\varepsilon_{\gamma}(E_{\gamma})\cdot
W(\theta))
\end{equation}
where $N_{\alpha-\gamma}(\theta,E_{\gamma})$ is the total number
of $\alpha-\gamma$ coincidences during the measurement time
$\Delta T_{meas}$, in the intervals of photon energies
$E_{\gamma}\pm\Delta E_{\gamma}/2$ and angles
$\Omega_{\gamma}\pm\Delta\Omega_{\gamma}/2$, $n_{\alpha}$ is the
intensity of particles in the $\alpha$-detector and
$\varepsilon_{\gamma}(E_{\gamma})$ is the absolute efficiency of
the $\gamma$-detector. Therefore, the total probability of the
bremsstrahlung emission is
\begin{equation}
dP/dE_{\gamma}=4\pi\cdot d^{2}P/dE_{\gamma}d\Omega_{\gamma} .
\end{equation}

The check on the  experimental data has been made by measuring the
coincidences between the $\alpha$-particles emitted to the first
excited level of the daughter nuclei $^{222}$Rn, $^{218}{\rm Po}$
and $^{210}{\rm Pb}$, corresponding to $\gamma$-rays with energies
186, 510 and 800 keV\cite{Firestone96}. The angular correlation
function W($\theta$), in the case of $\gamma$-ray E2-transitions,
can be presented as\cite{Ferguson69,Camp69}
\begin{equation}
W(\theta)=1+A_{2}\cdot Q_{2}\cdot P_{2}(cos\theta)+A_{4}\cdot
Q_{4}\cdot P_{4}(cos\theta),
\end{equation}
where  $A_{2}$ = 5/7, $A_{4}$= -12/7, $Q_{2,4}$ =
$Q_{2,4}^{\alpha}\cdot Q_{2,4}^{\gamma}$, $Q_{4}^{\alpha}$ = 0.68,
$Q_{4}^{\gamma}$ = 0.16 for $E_{\gamma}$ = 100 keV and
$Q_{4}^{\gamma}$ = 0.37 for $E_{\gamma}$ = 800 keV.

The bremsstrahlung spectra have been averaged over a photon energy
interval of  25 keV. The measured values of the photon emission probability
dP/dE$_{\gamma}$ due to the bremsstrahlung process accompanying
the $\alpha$-decay of $^{226}{\rm Ra}$ are shown by solid squares
in Fig. \ref{fig:226Ra_1}.

In the following Sec.\ref{sec.3} we calculate the bremsstrahlung spectrum accompanying the $\alpha$-decay of $^{226}{\rm Ra}$ and present, for a comparison, the above results for $^{226}{\rm Ra}$ (where is E$_{\alpha}$=4.8 MeV) with the ones presented in our previous paper~\cite{EPJA_2008} for  $^{214}{\rm Po}$ (where is E$_{\alpha}$=7.7 MeV).  In the theoretical description of the bremsstrahlung emission during the $\alpha$-decay there are some problems:  one is the choice of the realistic wave function of the $\alpha$-particle inside the nuclear potential\cite{Kurgalin01}. For example, the shape of the nuclear potential (the values of the nuclear radius R$_{n}$ and the deepness V$_{n}$ for a rectangular potential)  influences the slope of the wave function near the nuclear surface and therefore the conditions of tunneling through the Coulomb barrier. Other problems can be connected with the influence of the nuclear surface deformation and electron screening of the Coulomb barrier.

\section{Model, calculation and discussion
\label{sec.3}}

\subsection{Calculation method
\label{sec.3.1}}

 We define the
bremsstrahlung probability during the $\alpha$-decay of a nucleus
in terms of the transition matrix elements for the compound
quantum system ($\alpha$-particle and daughter nucleus) from its
state before photon emission (we name such a state as the
\emph{initial $i$-state}) into its state after the photon emission
(we name such a state as the \emph{final $f$-state}).
If it is possible to separate total wave function of
$\alpha$-particle (before and after photon emission) into radial
and spherical symmetric components (as in the approximation of the
spherically symmetric $\alpha$-decay), then one can find the
expression for the total bremsstrahlung probability with the
separation on the radial and angular components explicitly by an
analytical way.
Here, the radius defines the position of the particle with reduced
mass relatively to the center of mass. The angular components
contain all the detailed information about the directions of this
particle motion (with taking into account its tunneling) before
and after the photon emission and on the direction of the photon
emission.

According to Ref.~\refcite{Maydanyuk.2006.EPJA} we define the bremsstrahlung
probability as
\begin{equation}
  \displaystyle\frac{d P\, (w, \vartheta)}{dE_{\gamma}} =
  N_{0}\, k_{f}\, w \: \bigl| p(w, \vartheta) \bigr|^{2},
\label{eq.3.1.1}
\end{equation}
where
\begin{equation}
  N_{0} = \displaystyle\frac{Z_{\rm eff}^{2}\, e^{2}}{(2\pi)^{4}\, m},
\label{eq.3.1.2}
\end{equation}
\begin{equation}
\begin{array}{ccl}
  p(w, \vartheta) & = &
    -\sqrt{\displaystyle\frac{1}{3}}\, \cdot\,
    \displaystyle\sum\limits_{l=0}^{+\infty}
    i^{l} (-1)^{l} \: (2l+1) \: P_{l}(\cos{\vartheta})\, \cdot
    \displaystyle\sum\limits_{\mu = -1, 1} h_{\mu} J_{m_{f}}(l,w),
\end{array}
\label{eq.3.1.3}
\end{equation}
\begin{equation}
\begin{array}{ccc}
  h_{\pm} = \mp\displaystyle\frac{1}{\sqrt{2}} (1 \pm i), &
  k_{i,f} = \sqrt{2mE_{i,f}}, &
  w = E_{i} - E_{f}.
\end{array}
\label{eq.3.1.4}
\end{equation}
In (\ref{eq.3.1.3})  $J_{m_{f}}(l,w)$ is the radial integral independent on the angle $\vartheta$:
\begin{equation}
  J_{m_{f}}(l,w) =
    \int\limits^{+\infty}_{0} r^{2} \, R^{*}_{f}(r, E_{f}) \,
    \displaystyle\frac{\partial R_{i}(r, E_{i})} {\partial r} \, j_{l} (kr) \: dr.
\label{eq.3.1.5}
\end{equation}
In determination of the wave functions of the initial and final
states we use the selection rules for the quantum numbers $l$
and $m$:
\begin{equation}
\begin{array}{lll}
  \mbox{$i$-state before emission:} & l_{i} = 0, & m_{i} = 0;\\
  \mbox{$f$-state after emission: } & l_{f} = 1, & m_{f} = -\mu = \pm 1.
\end{array}
\label{eq.3.1.6}
\end{equation}
In (\ref{eq.3.1.5}) $j_{l}(kr)$ is the spherical Bessel function
of the order $l$, $R_{i}(r)$ and $R_{f}(r)$ are the
radial components of the total wave functions
$\psi_{i}(\mathbf{r})$ and $\psi_{f}(\mathbf{r})$ of the system in
the initial $i$- and final $f$-state, respectively.  For other notations and details see Ref. \cite{EPJA_2008}, in accordance with Refs.~\refcite{Maydanyuk.2006.EPJA,Maydanyuk.2003.PTP}.

To describe the interaction between the $\alpha$-particle and daughter nucleus ($A$, $Z$) we use the following potential~\cite{Denisov.2005.PHRVA}:
\begin{equation}
  V (r, \theta, l, Q) = v_{C} (r, \theta) + v_{N} (r, \theta, Q) + v_{l} (r)
\label{eq.3.2.1}
\end{equation}
where Coulomb $v_{C} (r, \theta)$, nuclear $v_{N} (r, \theta, Q)$ and centrifugal $v_{l} (r)$ components have such form:
\begin{equation}
\begin{array}{l}
  v_{C} (r, \theta) =
  \left\{
  \begin{array}{lcl}
    \displaystyle\frac{2 Z e^{2}} {r}
    \biggl(1 + \displaystyle\frac{3 R^{2}} {5 r^{2}} \beta_{2} Y_{20}(\theta) \biggr), &
    \mbox{for} & r \ge r_{m}, \\

    \displaystyle\frac{2 Z e^{2}} {r_{m}}
    \biggl\{
      \displaystyle\frac{3}{2} -
      \displaystyle\frac{r^{2}}{2r_{m}^{2}} +
      \displaystyle\frac{3 R^{2}} {5 r_{m}^{2}} \beta_{2} Y_{20}(\theta)
      \Bigl(2 - \displaystyle\frac{r^{3}}{r_{m}^{3}} \Bigr)
    \biggr\}, &
    \mbox{for} & r < r_{m}
  \end{array}
  \right.
\end{array}
\label{eq.3.2.2}
\end{equation}
and
\begin{equation}
\begin{array}{ll}
  v_{N} (r, \theta, Q) = \displaystyle\frac{V(A,Z,Q)} {1 + \exp{\displaystyle\frac{r-r_{m}(\theta)} {d}}}, &
  \hspace{10mm}
  v_{l} (r) = \displaystyle\frac{l\,(l+1)} {2mr^{2}}.
\end{array}
\label{eq.3.2.3}
\end{equation}
Here,  $Q$ is the $Q$-value for the $\alpha$-decay, $R$ is the radius of the daughter nucleus, $V(A,Z,Q,\theta)$ is the strength of the nuclear component; $r_{m}$ is the effective radius of the nuclear component, $d$ is the parameter of the diffuseness; $Y_{20}(\theta)$ is the spherical harmonic function of the second order, $\theta$ is the angle between the direction of the leaving $\alpha$-particle and the axis of the axial symmetry of the daughter nucleus; $\beta_{2}$ is the parameter of the quadruple deformation of the daughter nucleus. 
The parameters of the Coulomb and nuclear components are defined in Refs.~\refcite{EPJA_2008,Denisov.2005.PHRVA}.

In order to obtain the spectrum, we have to know wave functions in
the initial and final states. In the spherically symmetric
approximation one can rewrite the total wave functions by
separating the radial and angular components:
\begin{equation}
\begin{array}{cclcl}
  \varphi_{i}(r,\theta,\phi) & = &
    R_{i}(r) \: Y_{l_{i} m_{i}} (\theta, \phi) & = &
    \displaystyle\frac{\chi_{i}(r)}{r} \: Y_{l_{i} m_{i}} (\theta, \phi), \\
  \varphi_{f}(r,\theta,\phi) & = &
    R_{f}(r) \: Y_{l_{f} m_{f}} (\theta, \phi) & = &
    \displaystyle\frac{\chi_{f}(r)}{r} \: Y_{l_{f} m_{f}} (\theta, \phi).
\end{array}
\label{eq.3.3.1}
\end{equation}
We find the radial components $\chi_{i,f}(r)$ numerically on the
base of the given alpha-nucleus potential. Here, we use the
following boundary conditions: the $i$-state of the system before
the photon emission is a pure decaying state, and therefore for
its description we use the wave function for the $\alpha$-decay; after
the photon emission the state of the system is changed and it is
more convenient to use the wave function as the scattering of the
$\alpha$-particle by the daughter nucleus for the description of
the $f$-state. So, we impose the following boundary conditions on
the radial components $\chi_{i,f}(r)$:
\begin {equation}
\begin {array} {ll}
  \mbox {initial $i$-state:} & \chi_{i}(r \to +\infty) \to G(r)+iF(r), \\
  \mbox {final $f$-state:} & \chi _ {f} (r=0) = 0,
\end {array}
\label{eq.3.3.2}
\end {equation}
where $F$ and $G$ are the Coulomb functions as used in Refs.~\refcite{Maydanyuk.2006.EPJA,Maydanyuk.2003.PTP}.

\subsection{Bremsstrahlung components from different spacial regions, and interference term
\label{sec.3.4}}

It is interesting to estimate how much is the emission of photons from the tunneling region and the ones from the internal and external regions concerning the barrier versus the distance $r$.
Such a question was put for the first time by J.~Kasagi {\it et al.} in Ref.~\refcite{KasagiJPG} in the tentative to explain the difference between the experimental data  and the calculated spectrum for the $^{210}\mbox{Po}$ nucleus. 
Some later, a constructive analysis was proposed by N.~Takigawa {\it et al.} in Ref.~\refcite{Takigawa99} and, independently  E.~Tkalya in Ref.~\refcite{Tkalya99}, both giving the theoretical basis to work with the contributions of the photon emission from different regions and interference term by different approaches.
As we see and shall show further, the study of the photon emission from the different spacial regions allows one not only to analyze the difference between the experimental data and the calculated spectrum for $^{210}\mbox{Po}$, but gives a more important information about the studied process. For this reason, we shall develop a formalism of the photon emission from the different spacial regions in the framework of our model.

Let us define the bremsstrahlung probabilities of photons emitted from different spacial regions, characterized by the points $R_1$ and $R_2$ in Fig. \ref{barrier}.
According to Ref.~\refcite{Maydanyuk.2003.PTP}, we shall assume that the emission of photons from the internal spacial region (defined from $r=0$ up to  $r=R_{1}$) is very small in a comparison with the total photon emission, and therefore we can neglect it. In practical calculations of the spectra, we find the integral (\ref{eq.3.1.5}) by  a reasonable approximation taking an any finite value $R_{\rm max}$ into account instead of $r=+\infty$. By such reasons, we shall analyze the integral (\ref{eq.3.1.5}) only inside the spacial intervals from the point $R_{1}$ up to $R_{\rm max}$.

We separate this integral into two items:
\begin{equation}
  J_{m_{f}}(l,w) = J_{m_{f}}^{\rm (tun)}(l,w) + J_{m_{f}}^{\rm (ext)}(l,w)
\label{eq.3.4.1}
\end{equation}
where
\begin{equation}
\begin{array}{lcl}
  J_{m_{f}}^{\rm (tun)} (l,w) & = &
    \displaystyle\int\limits^{R_{2}}_{R_{1}} r^{2} \, R^{*}_{f}(r, E_{f}) \,
    \displaystyle\frac{\partial R_{i}(r, E_{i})} {\partial r} \, j_{l} (kr) \: dr, \\

  J_{m_{f}}^{\rm (ext)} (l,w) & = &
    \displaystyle\int\limits^{R_{\rm max}}_{R_{2}} r^{2} \, R^{*}_{f}(r, E_{f}) \,
    \displaystyle\frac{\partial R_{i}(r, E_{i})} {\partial r} \, j_{l} (kr) \: dr
\end{array}
\label{eq.3.4.2}
\end{equation}
where ($R_{1},R_{2}$) is the tunneling region, ($R_{2},R_{max}$) is the external one, and $E_i$ is the energy of the system in the initial $i$-state.
Then, the formula (\ref{eq.3.1.1}) of the bremsstrahlung probability is transformed into the following:
\begin{equation}
\begin{array}{cc}
 \displaystyle\frac{d P\, (w, \vartheta)}{dE_{\gamma}} =
 \displaystyle\frac{d P_{\rm tun}\, (w, \vartheta)}{dE_{\gamma}} +
 \displaystyle\frac{d P_{\rm ext}\, (w, \vartheta)}{dE_{\gamma}} +
 \displaystyle\frac{d P_{\rm interference}\, (w, \vartheta)}{dE_{\gamma}}.
\end{array}
\label{eq.3.4.3}
\end{equation}
where the three components $P_{\rm tun}$, $P_{\rm ext}$ and $P_{\rm interference}$ are defined through the finite integrals $J_{m_{f}}^{\rm (tun)}$ and $J_{m_{f}}^{\rm (ext)}$.

In particular, at $l=0$ (and assuming that the radial wave function $R_{f}(r)$ in the final $f$-state does not depend on quantum number $m_{f}$ at $l_{f}=1$ like as Coulomb functions) one can obtain:
\begin{equation}
\begin{array}{ccl}
  \displaystyle\frac{d P_{\rm tun}\, (w, \vartheta)}{dE_{\gamma}} & = &
    \displaystyle\frac{2}{3}\: N_{0}\, k_{f}\, w \; \Bigl| J^{\rm (tun)}\, (0,w) \Bigr|^{2},
\end{array}
\label{eq.3.4.4}
\end{equation}

\begin{equation}
\begin{array}{ccl}
  \displaystyle\frac{d P_{\rm ext}\, (w, \vartheta)}{dE_{\gamma}} & = &
    \displaystyle\frac{2}{3}\: N_{0}\, k_{f}\, w \; \Bigl| J^{\rm (ext)}\, (0,w) \Bigr|^{2},
\end{array}
\label{eq.3.4.5}
\end{equation}
\begin{equation}
\begin{array}{ccl}
  \displaystyle\frac{d P_{\rm interference}\, (w, \vartheta)}{dE_{\gamma}} & = &
    \displaystyle\frac{4}{3}\: N_{0}\, k_{f}\, w \;
    \Re \Bigl( J^{\rm (tun)}\, (0,w) \cdot J^{\rm (ext), *}\, (0,w) \Bigr).
\end{array}
\label{eq.3.4.6}
\end{equation}
Therefore, we affirm the following:
\begin{romanlist}[(ii)]
\item the total bremsstrahlung spectrum is not simply the summation of the direct (pure) probabilities from tunneling and external regions $P_{\rm tun}$ and $P_{\rm ext}$, but it also includes the interference term $P_{\rm interference}$ (see also Refs.~\refcite{Takigawa99,Tkalya99});
\item the probabilities $P_{\rm tun}$ and $P_{\rm ext}$ from tunneling and external regions are only positive, the interference term $P_{\rm interference}$ can be positive or negative.
\end{romanlist}

\subsection{Bremsstrahlung spectrum for $^{226}\mbox{\rm Ra}$
\label{sec.3.5}}

\begin{figure}[ht]
\vspace{1.5cm}
\centering
\resizebox{0.7\textwidth}{!}{\includegraphics{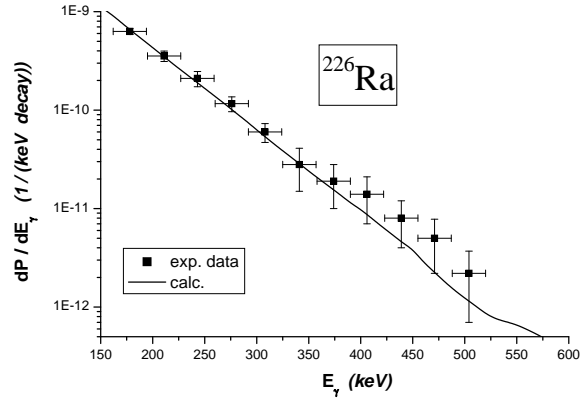}}
\vspace{-3.0cm} \caption{ The photon emission probability  $dP / dE_{\gamma}$ accompanying the $\alpha$-decay of $^{226}{\rm Ra}$. Full square are the experimental data, full line is the calculation for this nucleus.}
\label{fig:226Ra_1}
\end{figure}

We have applied the above described method to calculate the bremsstrahlung spectrum emitted during the $\alpha$-decay of the $^{226}\mbox{\rm Ra}$ nucleus.
The results are presented by the full line in Fig.~\ref{fig:226Ra_1} together with the experimental data (solid squares). In calculation we have used the approximation of the spherically symmetric $\alpha$-decay $l=0$ only in determination of $p(w, \vartheta)$ in (\ref{eq.3.1.3}), and the angle $\vartheta=90^{\circ}$ between the directions of the $\alpha$-particle motion (with possible tunneling) and the photon emission. The energy $E_{i}$ of the $\alpha$-particle in the initial $i$-state is $4.8$ MeV, according to Ref.~\refcite{Ndata}.

\begin{figure}[ht]
\vspace{2cm}
\centering
\resizebox{0.8\textwidth}{!}{\includegraphics{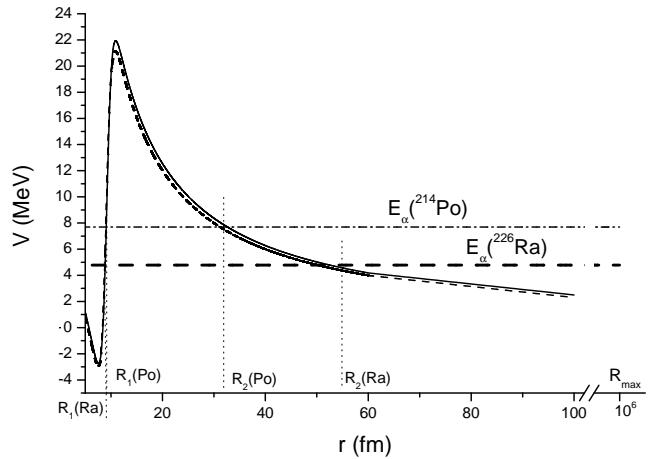}}
\vspace{-3.0cm} \caption{Alpha-nucleus potential $V$ versus r distance from the center of the decaying nucleus: $^{226}$Ra (full line), $^{214}$Po (dashed line). E$_{\alpha}$ represents the energy of the $\alpha$-decay for the $^{226}$Ra and  $^{214}$Po  nuclei, $R_1$ and $R_2$ are the points  delimiting  the tunneling region; $R_{max}$ is the maximum value of r considered in  calculation of the radial integral (\ref{eq.3.4.2}) instead of r=$+\infty$.}
\label{barrier}
\end{figure}

The approximation of the spherically symmetric $\alpha$-decay is used in the present paper for the following reasons (see also the analysis discussed in Ref.~\refcite{D'Arrigo93}). At the range of the nucleus surface and maximum value of the Coulomb barrier, the potential form changes more rapidly than for larger distances outside the nucleus surface and also inside the
nucleus (see Fig. \ref{barrier}).

Therefore at larger distances from the nucleus (and also
inside nucleus), where the changing of the potential is much
slower than in the surface region, the emission photon energy and influence of the deformation parameter of the nucleus both are
smaller than near the surface region. Considering that the
external region ($R_{2},R_{max}$) included between several fm and more large distances (much
\AA \, and further) is very wide, the photon emission
probability from such wide space is bigger than the one from the narrow
range of the nuclear surface region ($R_{1},R_{2}$). Therefore, the description of
the bremsstrahlung spectrum of Fig.~\ref{fig:226Ra_1} by the spherical-nucleus approximation
is rather good for lower photon energies (near about 100--300 keV
where the $\gamma$-emission probability $dP / dE_{\gamma}$ is
high). Of course, for photon energies larger than 400 keV it is
useful to take into account the deformation parameter of the
$^{226}{\rm Ra}$ nucleus ($\beta_2$=0.151)\cite{moller}. Such an improvement of the theory that takes into account the deformation parameter of the nucleus will be made in the next future because the upgrading of the model overcomes the aim of the present paper.

\begin{figure}[ht]
\vspace{2.83cm}
\centering
\resizebox{0.8\textwidth}{!}{\includegraphics{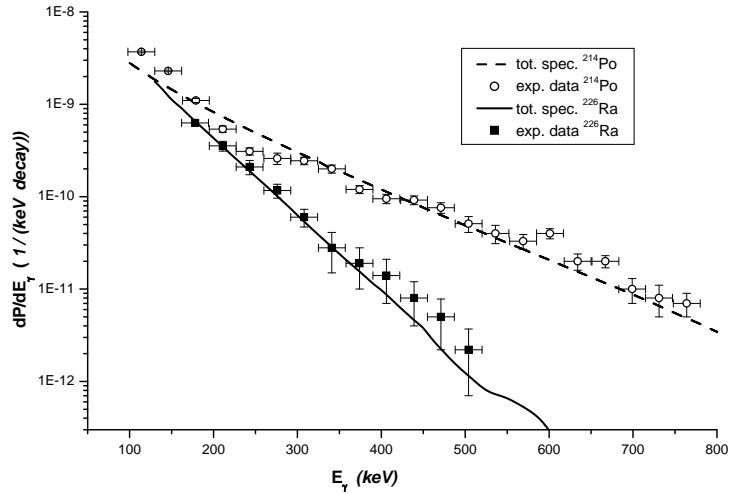}}
\vspace{-3.2cm}
\caption{
Calculation (full line) of the photon emission probability  $dP / dE_{\gamma}$ accompanying the $\alpha$-decay of $^{226}{\rm Ra}$ ($E_{\alpha} = 4.8$ MeV), and experimental data (full squares) $dP / dE_{\gamma}$ for $^{226}{\rm Ra}$ already presented in Fig.~\ref{fig:226Ra_1}.
For a comparison, we also include the experimental data (open circles)  and calculation (dashed line) obtained for the $\alpha$-decay of $^{214}{\rm Po}$  ($E_{\alpha} = 7.7$ MeV).
}
\label{fig:226Ra_214Po_full}
\end{figure}

In  Fig.~\ref{fig:226Ra_214Po_full} we report, for a comparison, the presented results for $^{226}{\rm Ra}$ together with the ones found for the $^{214}{\rm Po}$ nucleus\cite{EPJA_2008}. As Fig.~\ref{fig:226Ra_214Po_full} shows, both experimental and theoretical results of the photon emission probability obtained for the $\alpha$-decay of $^{226}{\rm Ra}$ are clearly lower than the ones obtained for $^{214}{\rm Po}$.  The difference between the two sets of data can be attributed to the different structure of the two nuclei, which affects the motion of the $\alpha$-particle inside the barrier. The ratio between the two sets of data of the photon emission probability $dP / dE_{\gamma}$ is strongly characterized by the different $\alpha$-decay energy for $^{214}{\rm Po}$ (E$_{\alpha}$=7.7 MeV)  and $^{226}{\rm Ra}$ (E$_{\alpha}$=4.8 MeV) concerning the shapes of the alpha-nucleus barriers for these nuclei.
In the $\alpha$-decay of $^{214}{\rm Po}$, the $\alpha$-particle with energy of 7.7 MeV  passes under the barrier in the upper part where the potential form changes more strongly emitting photons of high energies, while in the case of  $^{226}{\rm Ra}$ the $\alpha$-particle with energy of 4.8 MeV passes under the barrier in the lower part where the potential  changes partially more slowly emitting photons of lower energies.

\begin{figure}[ht]
\vspace{1.6cm}
\centering
\resizebox{1.00\textwidth}{!}
{
\includegraphics{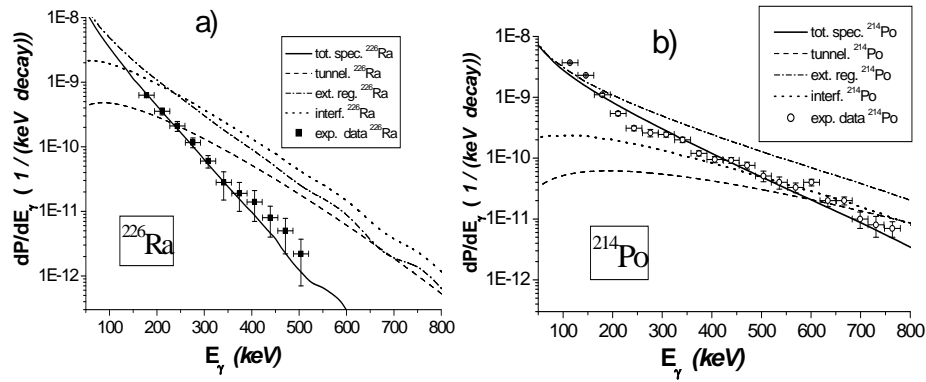}}
\vspace{-5.7cm}
\caption{
Spectra of photons emitted from the tunneling region (dashed line), from the external region (dash-dotted line), and the absolute value of contribution of the interference term (dotted line).
One can see for both nuclei the similar behavior of contributions from tunneling, external regions and interference term concerning the total spectrum.
}
\label{fig:226Ra_214Po}
\end{figure}

To explain the difference between the slopes of the bremsstrahlung spectra for the two considered nuclei, we formulate the following consideration: the slope of the bremsstrahlung spectrum is defined directly by the principal difference between the emission of photons during tunneling of the $\alpha$-particle and the emission of photons during its motion.
To understand why two nuclei have such different slopes of the bremsstrahlung spectra, we estimate how much the emission of photons during tunneling of the $\alpha$-particle differs from the emission of photons during further motion of this $\alpha$-particle in the external region.
In the Fig.~\ref{fig:226Ra_214Po} we report the tunneling and external contributions of the $\gamma$-emission accompanying the $\alpha$-decay of $^{214}{\rm Po}$ and $^{226}{\rm Ra}$ nuclei and the interference terms which are calculated by (\ref{eq.3.4.4})--(\ref{eq.3.4.6}). To make the analysis clearer, here we present the different contributions and interference terms for each nucleus separately. From calculation represented in this figure we establish that:

\begin{romanlist}[(ii)]
\item for both nuclei, the interference term has negative values in the whole energy region of the emitted photons and plays a destructive role in addition to the contributions from tunneling and external regions in  the forming total spectrum;
\item the calculated total emission probability (full line) for  $^{226}{\rm Ra}$ is always smaller than the one for $^{214}{\rm Po}$, and the external emission probability (dash-dotted line) for $^{226}{\rm Ra}$ is smaller than the one for $^{214}{\rm Po}$, for photon energies about E$_{\gamma}>$ 120 keV.
\end{romanlist}

%
Since the $\alpha$-particle energies for the decaying $^{214}{\rm Po}$ and $^{226}{\rm Ra}$ nuclei are 7.7 MeV and 4.8 MeV, respectively, the tunneling region ($R_1,R_2$) for $^{226}{\rm Ra}$  is  longer than for the case of $^{214}{\rm Po}$ (see Fig.~\ref{barrier}). Therefore, the bremsstrahlung photons are emitting through a greater distance under the barrier of the $^{226}{\rm Ra}$ nucleus than for $^{214}{\rm Po}$. For this reason, the relative contribution of the photon emission from the tunneling region into the total spectrum for $^{226}{\rm Ra}$ is larger than for $^{214}{\rm Po}$, at least up to about E$_{\gamma}$=450 keV (see the dashed  lines in panels a) and b) of Fig.~\ref{fig:226Ra_214Po}). In our results, the relative contribution of the photon emission from the external region ($R_2,R_{max}$) into the total spectrum for $^{226}{\rm Ra}$ is smaller than for $^{214}{\rm Po}$ (see dash-dotted lines in the cited panels a) and b)), because ($R_2,R_{max}$) is more large for  $^{214}{\rm Po}$ (see Fig. \ref{barrier}),  but such a contribution from the external region overcomes the total bremsstrahlung spectrum and  experimental data for both cases of the $^{226}{\rm Ra}$  and $^{214}{\rm Po}$ nuclei. This result clearly confirms that without the appreciable contribution of the photon emission during tunneling and the interference term (negative for these cases), it is impossible that the  total bremsstrahlung emission can reach and agree with the experimental results.  If the photon emission during tunneling of $\alpha$-particle is smaller than the photon emission during the external motion of this $\alpha$-particle, then we have to obtain smaller total bremsstrahlung spectrum for $^{226}{\rm Ra}$ than the spectrum for $^{214}{\rm Po}$ (at all photon energies). Figures ~\ref{fig:226Ra_214Po_full} and \ref{fig:226Ra_214Po} show  this effect and confirm (theoretically and experimentally) our results and considerations.
%

The smaller values of the calculated total emission probability for $^{226}{\rm Ra}$ than the one for $^{214}{\rm Po}$ (and the smaller values of the external emission probability for $^{226}{\rm Ra}$ than the one for $^{214}{\rm Po}$ can be explained by a consequence of the fact that outside the barrier the Coulomb field (and its derivative respect to $r$), that acts on the $\alpha$-particle, in the case of $^{226}{\rm Ra}$ is smaller than in the case of  $^{214}{\rm Po}$ because the external wide region results for the $^{214}{\rm Po}$ nucleus larger than for $^{226}{\rm Ra}$ and therefore the $\gamma$-emission probability for the $^{214}{\rm Po}$ nucleus is bigger.
\section{Conclusions}

We have obtained a good agreement between theory and experiment for the bremsstrahlung spectrum  of photons emitted during the $\alpha$-decay of the $^{226}{\rm Ra}$ nucleus, for E$_{\gamma}$ energies up to about 500 keV. We think that for photon energies higher than 400 keV it is useful to take into account the deformation parameter of the decaying nucleus in the model, in order  to reach a complete good agreement between data and calculation also at higher energies. We affirm that the photons with higher energies are emitted near the narrow region of the barrier where the potential form changes more rapidly, while the photons of lower energies are emitted at larger distances form the nucleus and in a very wide space where the changing of the potential is much slower. Therefore such a large wide space contributes with photons of low energies but with a higher emission probability. Moreover, we find that the slope of the bremsstrahlung spectrum accompanying the $\alpha$-decay of the $^{226}{\rm Ra}$ nucleus decreases more swiftly than the one registered for  $^{214}{\rm Po}$ because the $\alpha$-particle energy is lower (E$_{\alpha}$ = 4.8 MeV) for $^{226}{\rm Ra}$ than for $^{214}{\rm Po}$ (E$_{\alpha}$ = 7.7 MeV), and this plays an important role both on the motion of the $\alpha$-particle and in the determination of the photon emission probability in tunneling and external regions.
In both cases we find a destructive interference between the photons emitted from the two above-mentioned regions, and, for the first time, we find an important contribution of the bremsstrahlung photons emission during tunneling of the $\alpha$-particle in $^{226}$Ra, taking into account the realistic $\alpha$-nucleus potential in the model.

\section*{Acknowledgments}
G.~Giardina is grateful to the Fondazione Bonino-Pulejo (FBP) of Messina for the important  support received in the international collaboration between the Messina group and the Moscow State University (Russia) with the Institute for Nuclear Research of Kiev (Ukraine).
V.~S.~Olkhovsky and S.~P.~Maydanyuk thank the Dipartimento di Fisica dell'Universit\'a di Messina for  warm hospitality.


\end{document}